\documentstyle[preprint,aps,epsfig]{revtex}
\tightenlines
\newcommand{\lm}     {\lambda}
\newcommand{\M}     {{\mathcal M}}
\begin{document}
\preprint{\vbox{\baselineskip16pt
      \hbox{AS-ITP-2000-007}
      \hbox{SNUTP\hspace*{.2em}00-013}}
}
\title{
Scenario of light sterile neutrinos with a heavy tau neutrino
\\
in a supersymmetric model
}
\author{
Chun Liu$^{\:a}$ and Jeonghyeon Song$^{\:b}$
}
\vspace{1.5cm}
\address{
$^a$Institute of Theoretical Physics, Chinese Academy of Sciences\\
PO Box 2735, Beijing 100080, China\\
$^b$Department of Physics, Seoul National University,\\
Seoul 151-742, Korea
}
\maketitle
\thispagestyle{empty}
\setcounter{page}{1}
\begin{abstract}
Three light sterile neutrinos ($\nu_e^s$, $\nu_\mu^s$ and $\nu_\tau^s$) are
introduced to accommodate all the available neutrino data: the atmospheric 
neutrino anomaly is explained by $\nu_\mu-\nu_\mu^s$ oscillation with
maximal mixing; the solar one is due to $\nu_e-\nu_e^s$ oscillation of small
angle Mikheyev-Smirnov-Wolfenstein type; the Liquid Scintillation Neutrino
Detector data is from $\nu_e -\nu_\mu$ oscillations, so that the neutrinos
can be the hot component of the dark matter.  The big bang nucleosynthesis
constraint is satisfied by taking the tau neutrino to be $10$ MeV heavy.
The $\nu_\tau$ decay is discussed in a model of gauge mediated supersymmetry
breaking.  The decay mode $\nu_\tau\to\tilde{G}\gamma$ with $\tilde{G}$
being the gravitino is proposed.  The $\nu_\tau$ has a rather long lifetime
$\sim 10^3-10^{13}$ sec.  Its implication to the Gamma-ray Burst is
discussed.\\


\end{abstract}


\newpage
\section{Introduction}

Various experiments have provided growing evidence that neutrinos are
massive.  First the recent data on the atmospheric neutrinos from
Super-Kamiokande experiment has shown neutrino oscillations\cite{superk}.
It implies that the $\mu$-type neutrino has a maximal mixing with other
neutrino $x$ ($x\neq e$), and the mass squared difference is
$\Delta m_{\mu x}^2\simeq 3\times 10^{-3}$ eV$^2$.  The solar neutrino
deficit problem can be explained by either the Mikheyev-Smirnov-Wolfenstein
(MSW) solution \cite{msw} or the vacuum oscillation \cite{vac}.  The MSW
solution allows two sets of parameters:
$\Delta m_{ey}^2\simeq 5 \times 10^{-6}$ eV$^2$ with
$\sin^2 2 \theta_{ey} \simeq 6 \times 10^{-3}$, and
$\Delta m_{ey}^2\simeq 2 \times 10^{-5}$ eV$^2$ with
$\sin^2 2 \theta_{ey} \simeq 0.8$. The vacuum oscillation solution is
$\Delta m_{ey}^2\simeq 8 \times 10^{-11}$ eV$^2$ with
$\sin^2 2 \theta_{ey} \simeq 0.8$ 
\footnote {Recently the low mass solution has been revived which is 
$\Delta m_{ey}^2\simeq 10^{-7}$ eV$^2$ with large mixing \cite{low}.}.  
If the direct observation of
$\bar{\nu}_\mu - \bar{\nu}_e$ oscillations at the Liquid Scintillation
Neutrino Detector (LSND) experiment \cite{lsnd} is considered, relevant
parameters are $\Delta m_{e\mu}^2\simeq 1$ eV$^2$ and
$\sin^2(2\theta_{e\mu})\simeq 10^{-2}$.

In addition, astrophysics and cosmology also give us some information for
neutrino masses.  Various measurements support the existence of the dark 
matter.  One scenario for the dark matter is that not all of the 
cosmological dark matter is cold, there is some hot component \cite{COBE}.  
Canonical candidate for the hot dark matter (HDM) is the neutrino with mass 
at the scale of electron Volt \cite{COBE}.  If massive neutrinos are stable, 
their masses are limited to be less than few tens of eV from the standard 
cosmology, in order to avoid the over-closure of the Universe \cite{cos1}.

Non-vanishing neutrino mass might be the first discovery of the physics
beyond the standard model (SM).  The mass pattern of neutrinos is valuable
information for the exploration of the physics related to the flavor puzzle
in the SM.  Theoretically, several mechanisms have been suggested to
accommodate massive neutrinos.  One of the most popular scenarios is the
seesaw mechanism which naturally explains the smallness of neutrino mass by
introducing heavy right-handed neutrinos \cite{seesaw}.  Another example is
the supersymmetric extension of the SM with $R$-parity violation.  The
trilinear lepton number violating interactions induce small neutrino masses
at loop level \cite{rparity}.  However, the large mixing of neutrinos is not
expected from our experiences of the other SM fermions.

Phenomenological analyses suggest the existence of light sterile neutrinos.  
As least one light sterile neutrino is necessary to explain all the three
different scales of $\Delta m^2$ \cite{four}.
It is interesting and more natural that there are $three$ light sterile
neutrinos $\nu_i^s$ ($i= e,\mu,\tau$).  Then the first two generations can
fully explain all the above neutrino data.  In this framework, the neutrino
mixing pattern is uniquely fixed.  The atmospheric and solar neutrino
anomalies are due to the $\nu_\mu-\nu_\mu^s$ and $\nu_e-\nu_e^s$
oscillations, respectively.  The large mixing between $\nu_\mu$ and
$\nu_\mu^s$ can be understood naturally.  And the LSND evidence is
attributed to the $\nu_e-\nu_\mu$ mixing.  At least one of the neutrino
pairs should have mass around few eV, implying that neutrinos can be the hot
component of the dark matter.  This scenario was studied before
\cite{10,11}.  In this paper we reconsider it in a different theoretical
background.

In this scheme, more physics can be addressed.  The third generation
neutrinos have a separate story.  The above requirement for the first two
generations leaves a wider room for the $\nu_\tau$.  The $\nu_\tau$ can be
either ordinary with a mass smaller than $10$ eV, or exotic, for instance,
with $m_{\nu_\tau}\simeq (1-10)$ MeV \cite{10}.  The physics of
tau-neutrinos is influenced by the astrophysical constraint from the big
bang nucleosynthesis (BBN): the number of light neutrino species in thermal
equilibrium at the BBN era is limited to $N_\nu\leq 4.2$ \cite{12}.  It
should be seriously examined in a scenario with more than four light
neutrinos.  It is to be noted that the small mixing angle for the
$\nu_e-\nu_e^s$ oscillations causes only the active neutrinos to contribute
at the BBN era\cite{BBN1}.  The oscillation time is too long to make the
$\nu_e^s$ in thermal equilibrium.  In fact, the analyses for the solar
neutrino data have shown that the only viable pattern is the small angle MSW
type \cite{13}.  Therefore the first two generations contribute a factor
three to $N_\nu$.  As for the $\nu_\tau$ and $\nu_\tau^s$, at least one of
them should decouple at the time of BBN.  If $\nu_\tau$ is light
($m_{\nu_\tau}\leq 10$ eV), the sterile tau neutrinos must decouple from
other particles in the primeval plasma.  This is equivalent to the case of
introducing just two sterile neutrinos \cite{14}.  On the other hand, we may
consider $m_{\nu_\tau}\geq 1$ MeV so that $\nu_\tau$ decouples \cite{10}.
We are interested in this latter situation which has alternative theoretical
motivation \cite{15}.  In addition, the economy of the first two generation
oscillation scenario becomes more meaningful.

It is a theoretically challenging problem to construct a spectrum of very
light sterile neutrinos.  Several ideas have been suggested.  They are
related to extra symmetry \cite{A}, mirror world \cite{B}, axino \cite{C},
modulino \cite{D}, gravitino \cite{E}, string theory \cite{F}, gauge
mediated supersymmetry breaking (GMSB) \cite{G}, U(2) symmetry \cite{H},
seesaw mechanism \cite{I}, compositeness \cite{J}, top-flavor model \cite{K}
and extra dimensions \cite{L}.

This paper is organized as follows.  In the next section, we present a mass
matrix which is consistent with all the available neutrino data.  The
naturalness of the assumption is discussed.  In Sect.~III, the decay of the
tau neutrino is studied in a supersymmetric model.  The decay mode
$\nu_\tau\to\tilde{G}\gamma$ is proposed.  Further discussions including the
gamma-ray burst (GRB) are made in Sect.~IV.  The summary is given in the
last section.

\section{Neutrino Mass Matrix}

We assume that there are three light sterile neutrinos.  For each generation
of fermions, there is an associated sterile neutrino.  As will be seen in
the following, the atmospheric and LSND data will be accommodated naturally.
And the tau neutrino can be as heavy as ($1-10$) MeV without contradicting
experiments.  The full $6\times6$ neutrino mass matrix $\M$ consists of
three parts.  They are $3\times3$ active Majorana neutrino mass matrix
$M^{\rm active}$, $3\times3$ Dirac mass matrix $M^{\rm Dirac}$ which stands
for the mixing between the active and sterile neutrinos, and $3\times3$
Majorana mass matrix of the sterile neutrinos, respectively,
\begin{equation}
\label{1}
\M = \left(\begin{array}{cc}
M^{\rm active}&M^{\rm Dirac}\\
M^{\rm Dirac} &M^{\rm sterile}\\
\end{array}
\right)
\,.
\end{equation}

For the $M^{\rm active}$, we assume that except for the $\nu_\tau\nu_\tau$
component it originates from the ordinary seesaw mechanism \cite{seesaw};
and the $\nu_\tau\nu_\tau$ component is about ($1-10$) MeV.  The ordinary
seesaw mechanism suppresses neutrino masses to be $v^2/M_{\rm GUT}$, where
$v$ is the electroweak scale and $M_{\rm GUT}$ the grand unification (GUT)
scale so that $v^2/M_{\rm GUT}\sim 10^{-3}$ eV.  This is natural because
there are certain evidences  that the new physics beyond the SM is the GUT
\cite{a}. The introduction of the sterile neutrinos does not affect the
gauge coupling unification.  Furthermore, as can be seen below, this
$10^{-3}$ eV mass just fits the atmospheric neutrino data.  In other words,
the neutrino mass itself is also indirect evidence of the GUT.  Note that
even without the heavy right-handed neutrinos, the seesaw mechanism still
works in the GUT scenario.  It is from the general argument based on the
effective field theoretical point of view.  The MeV tau-neutrino has been
discussed in Ref. \cite{15}.  We introduce it here to avoid the BBN
constraint and to make the two generation oscillation scenario more
economical.

For the $M^{\rm Dirac}$, we adopt that it is proportional to the up-type
quark masses.  This similarity is understandable from the GUT point of view, 
if the sterile neutrinos are regarded as right-handed neutrinos.  It is
remarkable that this simple proportionality can be compatible with all the
current neutrino data.  In this quark mass matrix, the up quark is taken to
be $10^{-4} m_c$.  The up quark mass has not yet been determined
experimentally.  Even a massless up quark is allowed \cite{CP,c}.  

For the $M^{\rm sterile}$, we simply assume that it is approximately a
vanishing matrix.  This assumption makes the point of light sterile
neutrinos.  It could be relaxed to certain extent that the $M^{\rm sterile}$
plays no other roles in understanding the neutrino experiments. 

Since the main understanding of the neutrinos does not involve
intergenerational flavor mixing, without losing generality, we further
assume that all these mass matrices are diagonal.  We make a remark here
that the other zero entries of the Majorana mass matrices should be regarded
as $\sim 10^{-5}$ eV in general.  This is inevitable from the consideration
of quantum gravity \cite{d}.  However, such a number is not essential for
the understanding of the neutrinos in this framework.

Therefore the neutrino mass matrix $\M$ takes the following form,
\begin{equation}
\label{2}
\M = \left(\begin{array}{cccccc}
\displaystyle O\left(\frac{v^2}{M_{\rm GUT}}\right)&0&0&\lm m_u&0       &0\\
0&\displaystyle O\left(\frac{v^2}{M_{\rm GUT}}\right)&0&0      &\lm m_c &0\\
0&0                                       &10~{\rm MeV}&0      &0 &\lm m_t\\
\lm m_u                                            &0&0&0      &0       &0\\
0&\lm m_c                                            &0&0      &0       &0\\
0&0                                            &\lm m_t&0      &0       &0\\
\end{array}
\right)
\,,
\end{equation}
where $\lm$ is a constant.

Here starts numerical analyses.  As has been mentioned, $v^2/M_{\rm GUT}$ is
taken to be $3\times 10^{-3}$ eV.  We take $\lm m_c\simeq 1$ eV.  It is easy
to see that $\nu_\mu$ and $\nu_\mu^s$ have an almost maximal mixing with
$\Delta m^2\simeq 3\times 10^{-3}$ eV$^2$.  This explains the atmospheric
neutrino data.  The mixing between $\nu_e$ and $\nu_e^s$ is very small:
$\sin ^22\theta\sim [2\times 10^{-4}/(3\times 10^{-3})]^2\simeq 4\times
10^{-3}$ with $\Delta m^2\simeq 10^{-5}$ eV$^2$.  This is the small angle
MSW solution for the solar neutrino data.  The Dirac mass of tau neutrino is
$100$ eV.  This leads to no observable physical consequences of sterile tau
neutrinos because the Majorana mass of $\nu_\tau$ is about $10$ MeV which
makes the $\nu_\tau-\nu_\tau^s$ mixing to be at $\sim O(10^{-5})$.  As
discussed before, this small mixing precludes the sterile tau neutrinos from
contributing at the BBN era.  The LSND data of the $\nu_e-\nu_\mu$
oscillations is accounted by $\Delta m_{e\mu}^2 \simeq 1$ eV$^2$.  The
intergenerational mixings of the neutrinos are determined by the neutrino
mass matrix and the charged lepton mass matrix, which is too complicated to
be fixed in this framework.  Since the Dirac mass of the electron neutrino
is negligible, the $\nu_e-\nu_\mu$ mixing depends only on $M^{\rm active}$
and the charged lepton mass matrix.  In the case where $M^{\rm active}$ is
diagonal in certain weak basis, we expect that the charged lepton mass
matrix will fix the $\nu_e-\nu_\mu$ mixing, namely
$m_e/m_\mu\sim 5\times 10^{-3}$ which is consistent with the data.  The
$\nu_\mu$ and $\nu_\mu^s$ have masses around $1$ eV so that they can be the
HDM.  The $\nu_\tau$ cannot be stable due to cosmological considerations.

\section{The Decay of $\nu_\tau$}

In the previous section, phenomenology of three light sterile neutrinos is
discussed, and a heavy Majorana tau neutrino is necessarily assumed.  To
study the $10$ MeV tau neutrino and its decay, a theoretical model needs to
be specified.  First, this $\nu_\tau$ may decay via the ordinary weak
interaction, namely into $e^+e^-\nu_e$ through the $W$-boson exchange.
Its lifetime is
\begin{equation}
\label{3}
\tau_{\nu_\tau}\simeq \frac{192\pi^3}{G_F^2m_{\nu_\tau}^5|V_{e\tau}|^2}
\simeq 0.3\times \frac{1}{(m_{\nu_\tau}/10 {\rm MeV})^5 |V_{e\tau}|^2}~s
\,.
\end{equation}
Since the neutrino mass matrix leads to a very small mixing, only the
charged lepton mass matrix will determine $V_{e\tau}$.  It is generally
expected that $|V_{e\tau}|^2\simeq m_e/m_\tau\simeq 3\times 10^{-4}$ which
results in $\tau_{\nu_\tau}\simeq 10^3$ sec.  However this decay mode is
disfavored by the observation of the Supernova 1987A \cite{32a}.  Even if
this decay is marginally viable by assuming smaller $V_{e\tau}$, something 
more about the $\nu_\tau$ decay can be addressed.

There is an alternative decay channel which is irrelevant to the $e-\tau$
mixing.  It is inherent in the model of the ($1-10$) MeV $\nu_\tau$.  In
Ref. \cite{liu}, a supersymmetric model for the charged lepton masses is 
suggested.  In this model, because of family symmetries, $\tau$ lepton gets
mass from the nonvanishing vacuum expectation value (VEV) of the Higgs
field.  The muon gets mass only due to the sneutrino VEV which violates the
family symmetries.  The lepton number violation is allowed in the model.
The sneutrino VEV, which generates a $10$ MeV mass to the tau-neutrino, is
as large as ($5-10$) GeV, resulting in a mixing between the ordinary
neutralinos and the $\nu_\tau$ \cite{15}.  This model adopts the framework
of the GMSB \cite{A1}.  The $\nu_\tau$ in fact is the lightest neutralino 
(except for the gravitino) and the $\tau$ the lightest chargino.  Therefore
$\nu_\tau$ can decay to $\tilde{G}\gamma$ where the gravitino has mass of
few eV's.  From the neutralino mass matrix in Ref.~\cite{15}, it can be seen
that $\nu_\tau$ mixes with Bino at the level of
$m_{\tau H}/M_{\tilde{\gamma}}\sim 10^{-3}$ where $m_{\tau H}$ is the mixing
mass of the neutrino and neutralinos.  In ordinary GMSB models \cite{A1} the 
next-to-lightest neutralino, photino $\tilde{\gamma}$, decays to $\tilde{G}$ 
and $\gamma$.  The relevant matrix element 
$\langle\tilde{G}\gamma|\tilde{\gamma}\rangle$ can be written in the same way 
as in current algebra 
\begin{equation}
\langle\tilde{G}\gamma|\tilde{\gamma}\rangle=\frac{1}{F}
\langle\gamma|\partial_\mu\tilde{j}^\mu|\tilde{\gamma}\rangle
\,,
\end{equation}
where the supersymmetry current is 
\begin{equation}
\tilde{j}^\mu=F\gamma^\mu\tilde{G}+\sigma^{\mu\nu}\tilde{\gamma}F_{\mu\nu}
+...\,
\end{equation}
with $\sqrt{F}$ being the supersymmetry breaking scale $\sim 100$ TeV.  
From the second term in the above equation, the matrix element can be easily 
evaluated.  In our model, through considering the suppression factor 
$(m_{\tau H}/M_{\tilde{\gamma}})^2$, the $\nu_\tau\to\tilde{G}\gamma$ decay 
rate is then given by
\begin{equation}
\label{4}
\Gamma (\nu_\tau\to\tilde{G}\gamma)\simeq\left(
\frac{m_{\tau H}}{M_{\tilde{\gamma}}}\right)^2\cdot
\frac{\cos ^2\theta_Wm_{\nu_\tau}^5}{16\pi F^2}
\,.
\end{equation}
This is equivalent to a lifetime of $10^{13}$ sec, if we switch off the
$V_{e\tau}$.  Even though this lifetime is very long compared to the BBN
epoch, it is still short enough compared to the age of the Universe.

The BBN constraint for the number of light neutrino species is counted by
$\nu_e$, $\nu_\mu$ and $\nu_\mu^s$.  A lifetime of $10^3$ sec for $\nu_\tau$
is at the finishing stage of the BBN.  Such tau-neutrinos may be considered
as unstable in the BBN time.  With $\nu_e$ being the decay product, it is
probably helpful for the understanding of the Deuterium production and the 
structure formation in the cosmology \cite{B1}.  It is however more 
appropriate to take $\nu_\tau$ with lifetime of $\geq 10^3$ sec as stable in 
the BBN epoch \cite{B2}.  In this case, $\nu_\tau$ decouples in the BBN 
time.  And it must be Majorana neutrino \cite{B1,B2} as we have assumed.

\section{Discussions}

Several discussions are necessary about our scenario for neutrinos.  The
essential new ingredients of this scenario are the $10$ MeV Majorana
tau-neutrinos and the three light sterile neutrinos.  
A theoretical model of the $10$ MeV $\nu_\tau$ and its decay have been
discussed.  However, the lightness of the sterile neutrinos is still lack of
theoretical discussion.  To our knowledge, for more than one light sterile
neutrinos, there exists two attractive frameworks.  One is the compositeness
idea \cite{J}.  And the other comes from the mirror world \cite{B}.  They
are worthy to be studied further.  We introduce them phenomenologically in
this paper.

While the $R$-parity violating supersymmetry is used to generate the MeV
mass for $\nu_\tau$, it would be expected that the trilinear $R$-parity 
violating interactions contribute to the other entries of $M^{\rm active}$.  
However, this contribution is rather small.  It arises at one-loop level,
where a soft term of the relevant trilinear $R$-parity violating
interaction, $A$-term, is involved.  In the GMSB, $A$-terms are very small.
The contribution is in general smaller than $10^{-3}$ eV. 

Experimentally, our scenario can be tested in the very near future.  First,
the current laboratory mass limit for $\nu_\tau$ is $18.2$ MeV \cite{D1}.
It can be improved to be a few MeV in the near future, which could be a 
serious challenge to this scenario.

Second, future neutrino experiments will determine whether the atmospheric
neutrino deficit is due to $\nu_\mu-\nu_\mu^s$ or $\nu_\mu-\nu_\tau$ 
oscillation.  Although the present data favor the
$\nu_\mu-\nu_\tau$ oscillation, the $\nu_\mu-\nu_\mu^s$ oscillation cannot
be ruled out on the basis of the global fit to the full set of observable
\cite{E1}.  Recent Super-Kamiokande experiment claims that the 
$\nu_\mu\to\nu_\mu^s$ oscillation for the atmospheric neutrino anomaly is 
ruled out at $99\%$ C.L. \cite{50}.  However, it has been argued that more 
data and analyses are necessary before the conclusion is made \cite{51}. 
The long baseline neutrino experiments \cite{C0} will check the 
oscillation channel.  On the other hand, the oscillation channel for the 
solar neutrino deficit can be hopefully picked out by the new SNO 
observation \cite{future} together with previous experiments.
   
Third, the short baseline neutrino experiment BooNE will confirm whether
LSND anomaly is real.  
Even though the KARMEN \cite{karmen} group has reported that a large part of
the favored parameter region of LSND is excluded, full confirmation of the
LSND results still awaits future experiments.  For example, some of the LSND
solution space is preserved by a combined statistical analysis of both the
LSND evidence and the KARMEN exclusion \cite{Eitel}.

Cosmological and astrophysical aspects of this scenario deserve further
studies.  The implications on the BBN and HDM have been discussed in the 
previous sections.  For the decay mode $\nu_\tau\to\tilde{G}\gamma$, it 
might be the partial reason of the GRB \cite{F1}.  In the process of the 
stellar collapse, a huge number of tau neutrinos are emitted.  The decay 
product $\gamma$ is of few MeV or $100$ keV which is the typical energy 
scale of $\gamma$ in the GRB.  Suppose that the lifetime of $\nu_\tau$ is of 
order $10^{13}$ sec ( the $V_{e\tau}$ is negligible small $\leq 10^{-7}$), 
$10$ MeV $\nu_\tau$ emitted from the Supernova 1987A did not have enough 
time to decay when they arrived in the Earth \cite{grb}.  When the supernova 
occurs at cosmological distance ($\geq 10^{26}-10^{27}$ cm) from us, 
however, the GRB will be detected.  The ($1-10$) MeV $\nu_\tau$ may help our 
understanding of the cold dark matter (CDM) \cite{F0}.  As for the sterile 
neutrinos, it will be harmless for our numerical results to assume that 
$\nu_\tau^s$ has Majorana mass of thousand electron Volts.  Then this 
$\nu_\tau^s$ is also the candidate of the warm dark matter (WDM) \cite{G1}.  

On the other hand, if $\nu_\tau$ purely decays to $\tilde{G}\gamma$ with a 
lifetime $\sim 10^{13}$ sec, too much cosmic background radiation and 
matter density are caused \cite{cos1}.  For reducing the background 
radiation, one way maybe to prolong the $\nu_\tau$ lifetime from $10^{13}$ 
sec until to $10^{21}$ sec.  This can be achieved through adjusting the 
supersymmetry breaking scale (along with some other model parameters) up to 
$10^4$ TeV.  (In this case, the gravitino mass rises from few eV to 100 
keV.)  To avoid the matter density problem, we may need some non-standard 
cosmology.  For example, in a recent study \cite{50a}, it has been shown 
that if the largest temperature of the radiation era is as low as $0.7$ 
MeV, even a stable $10$ MeV tau neutrino is consistent with the cosmology.  

Within the framework of GMSB, the superpartners of the sterile neutrinos can
be very light because they are not involved in the gauge interactions.
However, their mixings with the ordinary sneutrinos which are typically
$100~{\rm GeV}-1$ TeV heavy, are very small.  So their lightness has no
influence on the physics of the BBN.  Meanwhile, these light scalar
particles can be also the candidate of the CDM \cite{49}.

Compared to the previous studies of MeV tau neutrinos, the distinguished 
feature of this scenario is that the ($1-10$) MeV tau neutrino is Majorana 
particle and its decay products do not contain Majoron.  Different from the
pseudo-Dirac neutrino scenarios\cite{11}, our explanation of the solar
neutrino deficit is due to small angle MSW solution.  The previous 
motivation was mainly to reduce the number of neutrino species in BBN era 
to be smaller than $3$, the present one is to keep that number just as $3$.  

\section{Summary}

Before summarizing the main points, let us list the dark matter candidates 
in this model.  Besides the $\nu_\mu$ and $\nu_\mu^s$ which have been taken 
as the HDM, there are many candidates for the CDM and the WDM.  They are  
($1-10$) MeV $\nu_\tau$'s, keV massive $\nu_\tau^s$'s, light sterile 
sneutrinos $\tilde{\nu}_i^s$ and gravitinos.  However, these are just 
possibilities.  Because the lifetime of $\nu_\tau$ has not been really 
fixed which still ranges from ($10^3-10^{13}$) sec (or longer); the 
heaviness of $\nu_\tau^s$ and $\tilde{\nu}_i^s$ are just assumed, which need 
to be studied from the theoretical mechanism for the lightness of sterile 
neutrinos.  Note that by taking the supersymmetry breaking scale as 
$\sim 100$ TeV, the gravitino is not the dark matter in simple GMSB models.  
However by raising the $\sqrt{F}$, keV $\tilde{G}$ can be the WDM 
(regardless the R-parity violation).  In addition, in the GMSB scenario, 
lightest stable baryons with mass of $\sim 100$ TeV in the hidden sector is 
also a candidate for the CDM \cite{A1}.  Among the five possibilities, the 
long-lived and decaying $\nu_\tau$ which is specific in this model, is 
reasonably favored to be the CDM in this model according to our discussion.  
For the stable $\nu_\tau$, the matter density was discussed in Ref. 
\cite{50a}.  It is expected that the same discussion can be made for the 
very long-lived $\nu_\tau$.  The requirement $\Omega_{\nu_\tau}h^2<1$ 
implies $m_{\nu_\tau}>3$ MeV for the reheating temperature being $0.7$ MeV.   

In summary, we have introduced three light sterile neutrinos $\nu_e^s$, 
$\nu_\mu^s$ and $\nu_\tau^s$ to accommodate all the available neutrino data.  
The atmospheric neutrino anomaly is explained by $\nu_\mu-\nu_\mu^s$
oscillation with maximal mixing and the solar one by $\nu_e-\nu_e^s$
oscillation of small angle MSW type.  The LSND data of the $\nu_e-\nu_\mu$
oscillations with $\Delta m_{e\mu}^2 \simeq 1$ eV$^2$ can be accommodated,
as providing canonical candidate of the HDM.  The BBN constraint is 
satisfied by taking the tau neutrino to be $10$ MeV heavy.  The $\nu_\tau$ 
decay has been discussed in a model of GMSB.  The decay mode 
$\nu_\tau\to\tilde{G}\gamma$ has been proposed.  The implication to the GRB 
has been pointed out.  Although the current data do not favor this 
oscillation scenario, it will be conclusively checked by atmospheric 
neutrino experiments, long baseline experiments and solar experiments in 
the near future.  

\acknowledgments

We would like to thank Carlo Giunti, Tan Lu for helpful discussions, and 
Xin-min Zhang for helpful discussions, drawing our attention to Ref. 
\cite{50a} and informing us that a similar scenario in the mirror world 
model with left-right symmetry was considered by R. N. Mohapatra and X.-M. 
Zhang.  This work was supported in part by the Fund of High Energy Physics 
Development of China, BK21 Program of the Ministry of Education of Korea 
and the National Natural Science Foundation of China.

\end{document}